\documentclass[twocolumn,preprintnumbers,floatfix,superscriptaddress,nofootinbib]{revtex4}
\pdfoutput=1 
\usepackage{CJKutf8} 
\usepackage{mathtools,slashed,mathrsfs,amsfonts}
\usepackage[caption=false]{subfig}
\usepackage{dcolumn}
\usepackage{multirow}
\usepackage{tabularx}
\usepackage{booktabs}
\usepackage{bm}
\usepackage{breqn}
\usepackage{comment}
\usepackage{setspace}
\usepackage[dvipsnames]{xcolor}
\usepackage[normalem]{ulem} 
\usepackage{enumerate}
\usepackage{siunitx}
\usepackage{multirow}
\usepackage{hyperref}

\definecolor{mypurple}{RGB}{164,64,214}

\newcommand\nn{\nonumber}
\newcommand\eea{\end{eqnarray}}
\newcommand\bea{\begin{eqnarray}}
\newcommand\ees{\end{split}}
\newcommand\bes{\begin{split}}

\def\l{\left(}
\def\r{\right)}

\begin{document}

%\vspace*{-30mm}

\title{Theta Dependence in the Presence of Massless Fermions
}

%%%%%%%%%%%%%%%%%%%%%%%%%%%%%%%%%%%%%%%%%%%%%%%%%%%%%%%

\author{Anson Hook}
\email{hook@umd.edu}
\author{Clayton Ristow}
\email{cristow@umd.edu}
\affiliation{Maryland Center for Fundamental Physics$,$ Department of Physics$,$ \\
University of Maryland$,$ College Park$,$ MD  20742$,$ U.S.A.}

%%%%%%%%%%%%%%%%%%%%%%%%%%%%%%%%%%%%%%%%%%%%%%%%%%%%%%%

\vspace*{1cm}

\begin{abstract} 

We show that if there are conserved flavor symmetries then some properties of a monopole can depend on $\theta$ even when a fermion is massless.  The quantized nature of global symmetries and the fractional nature of the Witten effect can lead to interesting structure.  Seen from another point of view, aside from possibly breaking baryon and lepton flavor symmetries (the Callan-Rubakov effect), monopole boundary conditions can also break the axial symmetry that otherwise could have been used to remove $\theta$ from the Lagrangian.  
%We study how the mass of the monopole acquires $\theta$ 
As an example, in a toy model, we calculate the $\theta$ dependence of the mass of the monopole and properties of the non-zero charge density surrounding the monopole.
%and the boost at which the Callan-Rubakov effect becomes unsuppressed.

\end{abstract}

\maketitle

\section{Introduction}

%1) Delta M_monopole vs theta.   2) Boost vs theta where you transition from chirality to lepton number violation.  3) charge density versus radius

The Lagrangian for electricity and magnetism has a topological term called the $\theta$ angle.  In the absence of any topologically non-trivial objects, this term is a total derivative and has no physical effect. Magnetic monopoles, being topologically non-trivial, acquire an electric charge $g \, \theta/2 \pi$ from this term, a phenomenon called the Witten effect~\cite{Witten:1979ey}.

In the presence of a light charged fermion, the Witten effect is more subtle.  Using a chiral rotation, $\theta$ can be moved into the mass term of the fermion. As a result, the Witten effect must be able to be seen from the point of view of the fermion ground state.  Solving the Dirac equation for a light fermion with a complex mass in the presence of a monopole~\cite{Wu:1976ge} and filling up the Dirac sea, it was found that the ground state carries fractional fermion number and an electric charge of $g  \theta/2 \pi$~\cite{Yamagishi:1982wp,Grossman:1983yf}.

However, when the mass of the fermion is exactly zero, the Witten effect is not present.  We therefore expect a non-zero monopole charge ($g \theta/2 \pi$) for a non-zero fermion mass, while for an exactly zero mass fermion, the charge of the monopole is exactly zero as well.  This discontinuity in the monopole's charge was rather disturbing for some time.  Eventually, it was realized that the charge density carried by the electron ground state is localized within a radius $r \sim 1/m_f$ of the monopole, where $m_f$ is the mass of the fermion.  Then, as the fermion becomes massless, the charge density spreads out over larger and larger distances. So, at any finite radius, one sees the charge density smoothly transitioning to zero resolving the confusion~\cite{Rubakov:1981rg,Rubakov:1982fp,Rubakov:1983sy,Callan:1982ah,Callan:1982au,Wilczek:1981dr}.

The vanishing of the Witten effect is often incorrectly stated as $\theta$ becoming unphysical in the presence of massless charged fermions.  This is {\it not} always correct in the presence of flavor symmetries.  
Let us take a $U(1)$ gauge theory with a light positively charged lepton and a heavier, also positively charged baryon.  
Imagine that the UV theory that gives rise to the 't Hooft-Polyakov monopole preserves a global $U(1)_{B-L}$ symmetry.  Now imagine taking a monopole and slowly turning on $\theta$.  Because the $U(1)_{B-L}$ charge commutes with the gauge symmetry, it does not change as $\theta$ is changed and the acquired $g \theta/2 \pi$ electric charge must be held equally by the baryon and lepton. 
Although the charge is held equally by the baryon and lepton vacua, the charge density around the monopole will be different for each due to their different masses. In particular, as the lepton mass goes to zero, a good portion of the leptonic charge is pushed off to infinity, while the baryonic charge remains concentrated at $m_B^{-1}$. This results in a non-trivial vacuum structure that depends on $\theta$, even as the lepton mass goes to zero. As a result, there are physical effects of $\theta$ even if the lepton is massless. 

%Because $U(1)_{B-L}$ is a good non-anomalous symmetry of the theory, it has quantized charge\footnote{The global symmetry must commute with the gauge group or else, as in $SU(5)$ GUTs, it can pick up a piece of the Witten effect.}.

Another way to see how $\theta$ remains physical when the lepton mass is zero comes from thinking about the Callan-Rubakov effect~\cite{Rubakov:1981rg,Rubakov:1982fp,Rubakov:1983sy,Callan:1982ah,Callan:1982au}. We will take the same IR theory discussed before which enjoys a global lepton number symmetry, $U(1)_L$, and a global baryon number symmetry, $U(1)_B$. Imagine a right-handed lepton incident on a monopole in a $J=0$ state. Conserving angular momentum, there are only two possible outgoing final states, a left-handed lepton or a left-handed baryon.  A right-handed outgoing lepton carries $J=1$ as the angular momentum stored in the gauge field changes sign as one goes from radially incoming to radially outgoing states.  As a result, scattering of a lepton off of a monopole either violates chirality or breaks baryon and lepton number.  Which of these is broken depends on the boundary conditions imposed on the lepton and baryon at the monopole, which in turn depends on the UV completion of the theory.

As made clear by the Callan Rubakov effect, monopole boundary conditions can break symmetries that might otherwise be present in the IR.  In particular, the chiral rotation that is used to remove $\theta$ from the Lagrangian can be broken by the boundary conditions imposed at the monopole.  If the boundary conditions violate chirality, then it is clear that the chiral rotation of the lepton that removes $\theta$ from the Lagrangian reinserts $\theta$ in the boundary condition at the monopole.  On the other hand, if the boundary conditions preserve chirality, then the chiral rotation can be used to remove $\theta$ from the theory entirely.

In what follows, we support our previous statements by explicitly calculating several physical effects associated with $\theta$ when $m_L = 0$, namely the $\theta$ dependence of the monopole mass and the non-zero charge density surrounding the monopole.

\section{The Model}
Our toy model is the one originally used by Callan when discussing the Callan Rubakov effect.  As such, we will only give a quick summary of the techniques used to solve the system and refer the reader to Ref.~\cite{Callan:1982au} for more details.  Our example is an $SU(2)$ gauge theory with $N_f = 2$ fermions and an adjoint scalar Higgs, $\Phi$. The theory also enjoys a $U(1)_{B-L}$ global symmetry.
\begin{center}
\begin{tabular}{c|c|c}
&$SU(2)$&$U(1)_{B-L}$ \\
\hline
$\Phi$&$Adj$&$0$ \\
$\psi = \begin{pmatrix}
b \\
l^c 
\end{pmatrix} $&$\Box$&1 \\
$\psi^c = \begin{pmatrix}
b^c \\
l 
\end{pmatrix}$&$\overline \Box$&-1
\end{tabular}
\end{center}
  
Through spontaneous symmetry breaking, $\Phi$ obtains a vev, $v$, and breaks $SU(2)$ down to $U(1)$ and the $\mathbb{Z}_2$ center of $SU(2)$, which results in a 't Hooft-Polyakov monopole~\cite{tHooft:1974kcl,Polyakov:1974ek} with magnetic charge $4\pi/g$. 
%Due to the unbroken $\mathbb{Z}_2$ center this is twice the expected minimal magnetic charge of the monopole.\CR{This is the same as saying that the minimum electric charge is $1/2$, correct?}  
The fermions obtain a mass from a term in the Lagrangian of the form
\bea
- \mathcal{L} \supset m \psi \psi^c + y \psi \Phi \psi^c = m_B b b^c + m_L l l^c, 
\eea
with $m_B = m + y \, v$ and $m_L = m - y \, v$.  Note that while our choice of language is inspired by the real world, our example is not meant to actually reproduce the real world but merely illustrate a point, e.g. our lepton has charge $1/2$ as opposed to $-1$.
This particular way of embedding the lepton and baryon into the UV $SU(2)$ theory imposes chirality violating boundary conditions.

There is an alternative UV completion that realizes the $B-L$ violating boundary condition.  The two UV $SU(2)$ doublet fermions would be constructed as $\psi_{b} = \begin{pmatrix}
b \\
b^c 
\end{pmatrix}$ and $\psi_{l} = \begin{pmatrix}
l \\
l^c 
\end{pmatrix}$ with mass terms of the form $y_B \psi_b \Phi \psi_b$ with analogous terms for the leptons.  If $m_L = 0$, one could remove $\theta$ from the UV theory by rotating $\psi_l$.  As $\theta$ really is unphysical in this particular UV completion, we will not discuss it any further.

Around a monopole, $\psi$ and $\psi^c$ can be decomposed into angular momentum states eigenstates. As the fermionic ground state is likely minimized by spherically symmetric solutions, we follow Callan and consider only the $J = 0$ state.  Restricting ourselves to this limit, the $J=0$ mode of each of the fermion doublets is written as a matrix of spin and gauge indices with the form
\bea
\label{Eq: J=0 Fermions}
\psi = \frac{g(r,t) + p(r,t) \hat r_i \sigma_i}{\sqrt{8 \pi r^2}} \sigma_y .
\eea
The monopole solution breaks UV rotations and the $SU(2)$ gauge group down to the diagonal.  There are two ways to get a rotationally invariant state: if the state is a singlet under both UV rotations and $SU(2)$, or if the state transforms under both symmetries but is a singlet when both are done at the same time. These two pieces are captured by the $g$ and $p$ terms respectively in Eq.~\ref{Eq: J=0 Fermions}. In the limit of an infinitely small monopole, exciting the $p(r,t)$ mode inside of the monopole core costs infinite energy as the angular momentum barrier reemerges implying a boundary condition at the origin with $p(0,t) = 0$.  This specification of the boundary conditions also involves gauge fixing as $g$ and $p$ rotate into each other under a $U(1)$ gauge transformation.

As a result of this simplification, our four-dimensional problem reduces to a two-dimensional problem complete with two 2D fermions built out of the two $SU(2)$ doublets. Famously, there is a duality between two-dimensional fermions and bosons~\cite{Mandelstam:1975hb,Coleman:1974bu}.  Very schematically, this duality proceeds as
\bea
\psi_{2D} = \begin{pmatrix}
b(r-t) \\
l^c(r+t)
\end{pmatrix} \sim \begin{pmatrix}
e^{i \phi(r-t)} \\
e^{i \phi(r+t)}
\end{pmatrix}
\eea
with a similar expression for the other doublet with a different scalar. In this basis of scalar fields, the mass terms are off-diagonal so as a final simplification, a canonical transformation is made on the scalar fields to diagonalize the mass terms. 
After integrating out the $U(1)$ gauge boson, the two bosons $\phi_B$ and $\phi_L$ have the action
\bea
\label{Eq: 2D Action}
S &=& \int_{r>0} dr \, dt\, \frac{1}{8 \pi} (\partial \phi_B)^2 + \frac{1}{8 \pi} (\partial \phi_L)^2 + \frac{\pi m_B^2}{16} \cos \phi_B  \nonumber \\
& &+ \frac{\pi m_L^2}{16} \cos \phi_L - \frac{g^2}{8 \pi r^2} \l\frac{\phi_B}{4 \pi} + \frac{\phi_L}{4 \pi} - \frac{\theta}{2 \pi}\r^2 .
\eea
Far from the monopole, this sine-Gordon theory has solitons and anti-solitons for both $\phi_B$ and $\phi_L$.
This bosonic theory is related to the original fermionic theory as a $\phi_B$ (anti) soliton is a (anti) baryon while a $\phi_L$ (anti) soliton is an (anti) lepton.  The mass terms are normalized such that the energy of a soliton at rest is equal to that of a fermion at rest\footnote{While we are really interested in the $m_L = 0$ limit, we will also be discussing $m_L > 0$ but small and taking the limit as it goes to zero.  Taking a positive mass avoids the annoying issue of charge fractionalization in this theory~\cite{Jackiw:1975fn}.}.    
Meanwhile, we study the minimal positively charged monopole, as opposed to the excited dyon states, by restricting $0 < \theta < \pi$, which will result in $\phi_{B,L}$ going to $0$ as $r \rightarrow \infty$. All of the physical effects of negatively charged monopoles can be easily obtained from their positively charged counterparts.

As emphasized before, boundary conditions are critical in analyzing the ground state.  Translating the $p(0,t) = 0$ boundary conditions into the bosonized language Callan obtained
\bea
\label{Eq: BC}
\phi_B(0) = \phi_L(0) \qquad \partial_r \phi_B(0) = - \partial_r \phi_L(0).
\eea 
The first of these boundary conditions combined with the asymptotic behavior of $\phi_{B,L}$ is what forces the conserved $B-L$ charge to be an integer.  
Amusingly, the second boundary condition is unnecessary as all of our minimum energy solutions obey $\partial_r \phi_B(0) = \partial_r \phi_L(0) = 0$.  
The electric charge density can be obtained using
\bea
\label{Eq: Charge Density}
4 \pi r^2 \rho(r) =- \frac{1}{4 \pi} \partial_r (\phi_B + \phi_L).
\eea
Given that $\phi$ asymptotes to $0$, the charge of any configuration can be read from
\bea
Q_B = \frac{\phi_B(0)}{4 \pi} \qquad Q_L = \frac{\phi_L(0)}{4 \pi} ,
\eea
where $Q_B$ and $Q_L$ are the charges carried by the baryon and lepton respectively.

%\medskip
%\noindent\uline{\textbf{Interpretation of Bosonization}} 
%\medskip

%\AH{added this section}

%We conclude this section with a discussion of the physical interpretation of bosonization.  Around a monopole there are a variety of condensates.
%\bea
%\langle b b^c \rangle \sim \frac{m_B}{r^2} &\qquad& \langle l l^c \rangle \sim \frac{m_L}{r^2} \\
%\langle b l^c \rangle = \langle l b^c \rangle \sim \frac{1}{r^3} &\qquad& r \lesssim \frac{1}{m_B} \\
%\langle b l^c \rangle = \langle l b^c \rangle \sim e^{-m_B r} &\qquad& r \gtrsim \frac{1}{m_B}.
%\eea
%The scalars $\phi_{L,B}$ are simply the phases of these condensates.  Thus when we later minimize the energy and solve for the ground state configuration of $\phi_{L,B}$ as a function of $\theta$, we are simply solving for the phases of these condensates as a function of distance from the monopole.
\begin{figure}[t]
 \centering
 \includegraphics[width=0.49\textwidth]{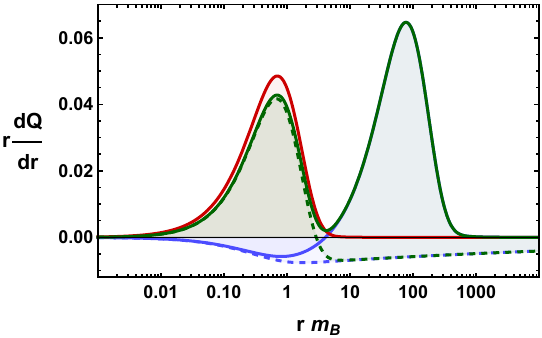}
 \caption{A picture of the baryon (red), lepton (blue), and total (green) differential charge densities as a function of log radius when $\theta = \pi/2$ and $\alpha = 1$.  In solid (dashed) lines, we have the charge density when $m_L = m_B/100$ ($m_L = 0$).   The baryon and lepton vacuums each carry charge equal to $g \theta/4 \pi$.  The baryonic charge of $g \theta/4 \pi$ is located roughly at distances $r \sim 1/m_B$.  The lepton proceeds to screen the charge deposited by the baryon by depositing negative charge at distances $r \sim 1/m_B$, albeit in an extremely inefficient manner.  At distances $r \sim 1/m_L$ the lepton then deposits a charge to bring its total charge to $g\theta/4\pi$.}
    \label{charge1}
\end{figure}

\section{Results}

\begin{figure}[t]
    \centering
    \includegraphics[width=0.49\textwidth]{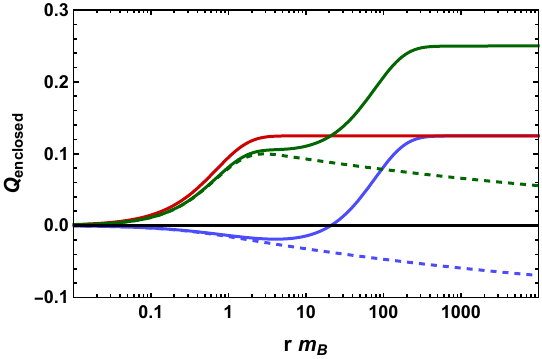}
    \caption{Same as Fig.~\ref{charge1} but instead plotting the total charge enclosed as a function of radius.}
    \label{charge2}
\end{figure}

We find the fermion ground state by solving the equations of motion, which can be found from Eq.~\ref{Eq: 2D Action} to be
\bea
\label{Eq: Eq of Mo}
\left ( \partial_t^2 - \partial_r^2 \right ) \phi_{B,L} &=& - \frac{g^2 (\phi_B + \phi_L - 2 \theta )}{16 \pi^2 r^2} \nonumber \\
& & - \frac{\pi^2 m_{B,L}^2}{4} \sin \phi_{B,L} .
\eea
The boundary conditions in Eq.~\ref{Eq: BC} combined with finite energy enforce $\phi_B(0) = \phi_L(0) = \theta$ fixing each fermion field to carry a charge $g\theta/4\pi$.  While we have numerically solved for the ground state of the monopole at various values of $\theta$ and $\alpha = g^2/4 \pi$, let us first discuss analytic approximations of the solution.  We are interested in the limit $m_B \gg m_L$. At large radii $(r\gg m_L^{-1})$, the solution for both $\phi_B$ and $\phi_L$ are found by balancing the electromagnetic term against the mass term in Eq.~\ref{Eq: Eq of Mo} yielding
\bea
\phi_{B,L}(r\gg m_L^{-1}  ) \approx \frac{2\theta\alpha}{\pi^3 m_{B,L}^2r^2}.
\eea
%\CR{If $\alpha/4\pi\gg 1$, the limits should be $m_L^{-1}\sqrt{\alpha/4\pi}\ll r$. Not sure if that's relevant since we don't really consider $\alpha/4\pi\gg 1$}
Interestingly the charge density does not fall exponentially at large $r$ but instead falls polynomially.  Meanwhile, if $r\ll\sqrt{\frac{\alpha}{4\pi}}m_L^{-1}$, the mass term for $\phi_L$ becomes irrelevant. Here one can set the $\phi_L$ mass term to zero and solve for $\phi_L$. $\phi_B$ can simultaneously be found again balancing the electromagnetic term with the $\phi_B$ mass term. The solutions in this region are then
\bea
\label{Eq: Asymptotic mL=0}
&\phi_L\l m_B^{-1} \ll r \ll m_L^{-1}\sqrt{\frac{\alpha}{4\pi}} \r - 2\theta \propto r^{\frac{1}{2} (1 - \sqrt{1 + \frac{\alpha}{\pi}} ) } \\
&\phi_B\l m_B^{-1} \ll r \ll m_L^{-1}\sqrt{\frac{\alpha}{4\pi}} \r\approx \frac{\alpha}{\pi^3m_B^2r^2}(2\theta-\phi_L(r))\nn.
\eea
We find good agreement when comparing these analytic approximations with the exact numerical results.

Our numerical solution for the scalar fields $\phi_{B,L}$ can be used to give the charge density or the total charge enclosed as a function of r.  We present our results in Fig.~\ref{charge1} and Fig.~\ref{charge2}.  As can be seen, the baryonic vacuum deposits its charge $g \theta/4\pi$ at a radius $r \sim 1/m_B$.  The leptonic vacuum reduces the electromagnetic energy by partially screening the baryon vacuums charge. However, since $m_L\ll m_B$, the charge density in the leptonic vacuum cannot jump as sharply as the baryonic vacuum and thus is unable to screen the entire charge.  Finally, at a radius $r \sim 1/m_L$, the lepton gives the Witten effect by depositing the necessary charge to bring its total charge to $g \theta/4\pi$.  

From Fig.~\ref{charge1} and Fig.~\ref{charge2}, it is clear that vacua's $\theta$ dependence does not vanish in the $m_L = 0$ limit.  Aside from simply looking at the $m_L  = 0$ lines on the plot, one can easily visually imagine the $m_L = 0$ limit by simply sitting at a finite radius and slowly letting the radius at which the extra charge in the leptonic vacuum is deposited approach infinity.\\ %At any finite radius, the remaining charge in the leptonic vacuum has not been depositied and the Witten effect has not taken place. At this finite radius one would observe some charge density due to the imperfect screening by the leptonic vacuum. In principle, one could probe this charge density experimentally. Despite having no total electric charge, the charge density surrounding the monopole is non-vanishing.

%\begin{figure}[t]
%    \centering
%    \includegraphics[width=0.49\textwidth]{Loop.pdf}
%    \caption{A picture of the ground state when $\theta = \pi/4$, $m_B = 1$ and $m_L = 0.01$.  As can be seen, $\theta$ is screened by a combination of the baryon and lepton.  The baryon and lepton vacuums each carry half of the theta angle.  The baryonic vacuum has structure on distances $r \sim 1/m_B$ while the lepton vacuum has structure on distances $r \sim 1/m_B$ and $r \sim 1/m_L$.
%    }
%    \label{ground}
%\end{figure}

\begin{figure}[t]
    \centering
    \includegraphics[width=0.49\textwidth]{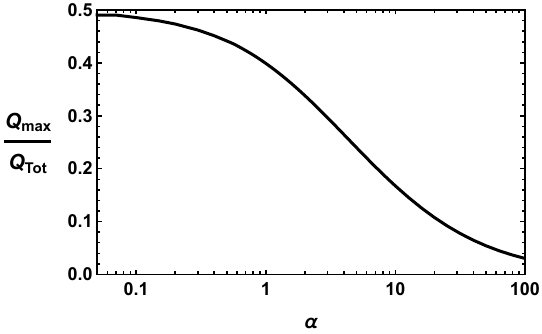}
    \caption{We take a monopole with $m_L = 0$ and calculate the largest the charge enclosed by a radius r gets, $Q_{\rm max}$.  We plot $Q_{\rm max}$ versus $\alpha$ normalizing $Q_{\rm max}$ by the total charge of the monopole when $m_L \ne 0$, $Q_{\rm tot}=g \theta/2 \pi$.  For large values of $\alpha$, the enclosed charge is never large.  For small values of $\alpha$, it quickly asymptotes to the charge held by the baryons.}
    \label{screen}
\end{figure}

%\smallskip
\noindent\uline{\textbf{Charge Density}} 
\medskip

One of the interesting observations from considering the limit where when $\theta \ne 0$ and $m_L = 0$, is that, despite the total charge in the vacuum being zero, the charge density in the vacuum is non-vanishing. This can easily be seen in Fig.~\ref{charge2} where, in the $m_L=0$ limit, the total charge enclosed at radius $r$, $Q_{\rm enclosed}(r)$ (the green dashed line), is non-zero even for large $r$. At a radius $r\sim m_B^{-1}$, $Q_{\rm enclosed}$ is maximized. Afterwards, the enclosed charge falls off as $r^{\frac{1}{2}(1-\sqrt{1+\alpha/\pi})}$, as can be seen from the analytic solution in Eq.~\ref{Eq: Asymptotic mL=0}. This power law falloff can be extremely slow, falling as $r^{-\alpha/4\pi}$ for $\alpha\ll1$.

The maximum enclosed charge, $Q_{\rm max}$, is proportional to $\theta$ and has a more complicated dependence on $\alpha$.  We plot this dependence as a function of $\alpha$ in Fig.~\ref{screen}.

When $\alpha$ is large, $Q_{\rm max}$ can be highly suppressed, while for $\alpha\sim1/137$, the screening of the charge by the lepton is very inefficient and the maximal charge is approximately the charge held by the baryons $g \theta/4 \pi$. This can be understood heuristically from the equations of motion. At $r\sim m_B^{-1}$, the electromagnetic term is $\sim \frac{\alpha m_B^2}{4\pi}$. In the $m_L=0$ limit, this fixes the rate at which $\phi_L$ can change to try and screen the charge from the baryon vacuum $\partial_r^2\phi_L\sim \frac{\alpha m_B^2}{4\pi}$. The baryon vacuum's rate of change is based on which of the two terms, electromagnetic term or mass term, is greater: $\partial_r^2\phi_B\sim m_B^2 \max(1,\frac{\alpha}{4\pi})$. So we can see in the $\alpha/4\pi\ll 1$ limit, $\partial_r^2\phi_L\ll \partial_r^2\phi_B$ and so the lepton vacuum does a poor job screening the baryon's charge since it cannot change fast enough to compensate. In the other limit $\alpha/4\pi\gg 1$, $\partial_r^2\phi_B\sim\partial_r^2 \phi_L$ and so the leptonic vacuum is able to effectively screen all of the charge in the baryonic vacuum.

\begin{figure}[t]
    \centering
    \includegraphics[width=0.49\textwidth]{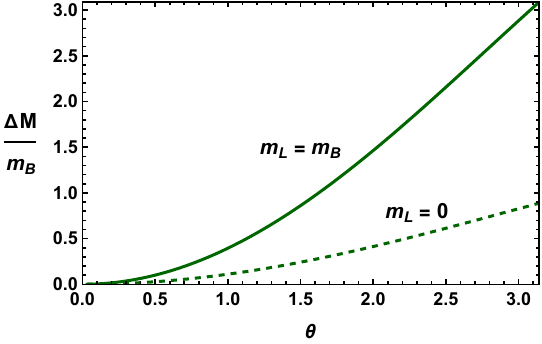}
    \caption{ The dependence of the monopole mass normalized to $m_B$ on $\theta$ when $m_L = m_B$ (solid line) and when $m_L = 0$ (dashed line).  In both cases we take $\alpha = 1$.}
    \label{mass}
\end{figure} 

\medskip
\noindent\uline{\textbf{Monopole Mass}} 
\medskip

Finally, it is clear that the solution at non-zero $\theta$ carries energy and will therefore contribute to the mass of the monpole.  As such, we can calculate the $\theta$ dependence of the monopole mass in the $m_L = 0$ limit.  This is shown in Fig.~\ref{mass}.  Dimensional analysis forces the $\theta$ dependence of the mass to be proportional to $m_B$.  Amusingly, we find that the dependence of the monopole mass on $\theta$ only changes at the $\mathcal{O}(1)$ level as the lepton mass goes from equal to the baryon mass to zero.  In both the equal mass and zero mass limit, the scaling of the mass is  $\Delta M \propto \theta^2$.   %The $\alpha$ dependence is complicated and even non-monotonic as $\alpha$ becomes larger than $\mathcal{O}(1)$.

%Finally, the Callan Rubakov effect has strong velocity dependence.  If a $J=0$ baryon is incident on a monopole, either a lepton or baryon can emerge.  At low velocities, the chirality violation due to the mass term is more important than the baryon number violating boundary conditions and the final state is a baryon.  Meanwhile at high velocities, the mass term is less important than the boundary conditions and a lepton emerges.  The boost at which one transitions from one limit to the other has been studied before, see e.g. Ref.~\cite{Dawson:1983cm}.

%We now study how this transition between final states depends on $\theta$.  We calculate the critical boost a which baryon scattering transitions from an outgoing baryon to an outgoing lepton and present our results in Fig.~\ref{Callan}.

\section{Conclusion}

In this article, we explored the $\theta$ dependence of various monopole related quantities in the presence of a massless fermion.  Contrary to what is often stated, despite the presence of a chiral rotation that can be used to rotate away $\theta$, $\theta$ can reappear in boundary conditions imposed at the core of a monopole, so that physical quantities can depend on $\theta$.  
We show that while the charge of the monopole is always zero in the massless fermion limit, the Witten effect is absent, and the charge density surrounding a monopole is non-zero and $\theta$ dependent.  Additionally, we showed how the mass of the monopole acquires $\theta$ dependence that is parametrically similar to the case of a massive fermion.  There are quite a number of ways in which the effects discussed in this letter might be important to various Beyond the Standard Model theories.  For example, the $SU(5)$ GUT monopole is the $N_f = 4$ version of our example~\cite{Callan:1982au,Dawson:1983cm} and has all of the properties discussed, e.g. $\theta$ dependence survives even if the massless down quark were utilized to solve the strong CP problem.  Due to the overlap between global symmetries and electric charge at low energies, acquiring fractional electric charge and integer global charge is even more complicated than our toy model and new exciting phenomena appear.  Further discussion along these lines would inevitably result in citations to papers written after the birth of the authors and as such will be left to future work.
%With all of these interesting $\theta$ dependent quantities, it is quite unfortunate that in the Standard Model, there are no neither massless fermions nor monopoles.  
%Perhaps we will be lucky and some of this interesting physics will manifest itself instead in a dark sector.

\section*{Acknowledgments}

%We thank xxx for useful discussions.  
AH and CR are supported in part by the NSF grant PHY-2210361 and the Maryland Center for Fundamental Physics.

\bibliography{biblio}{}

\bibliographystyle{JHEP}

\end{document}